\documentclass{aa}

\usepackage{subfig}

\usepackage{array} 
\usepackage{tabularx}
\usepackage{graphicx}
\usepackage[version=3]{mhchem}
\usepackage{txfonts}
\newcommand{\tablenotea}[1]{\parbox{6.6cm}{ \indent \footnotesize{\textsc{}~#1}}}

\usepackage{natbib}  
\bibpunct{(}{)}{;}{a}{}{,} 

\begin{document}
\title{The atmospheric chemistry of the warm Neptune GJ~3470b: influence of metallicity and temperature on the \ce{CH4}/CO ratio}
\titlerunning{The atmospheric chemistry of the warm Neptune GJ~3470b}
\authorrunning{Venot et al.}

\author{Olivia Venot\inst{1}, Marcelino Ag\'undez\inst{2,3}, Franck Selsis\inst{2,3}, Marcell Tessenyi\inst{4}, Nicolas Iro\inst{5}}

\institute{Instituut voor Sterrenkunde, Katholieke Universiteit Leuven, Celestijnenlaan 200D, 3001 Leuven, Belgium \and 
Univ. Bordeaux, LAB, UMR 5804, F-33270, Floirac, France \and 
CNRS, LAB, UMR 5804, F-33270, Floirac, France \and 
University College London, Department of Physics and Astronomy, Gower Street, London WC1E 6BT, UK \and 
Theoretical Meteorology group, Klimacampus, University of Hamburg, Grindelberg 5, 20144, Hamburg, Germany \\
\email{olivia.venot@ster.kuleuven.be}}

\date{Received; accepted}


\abstract
{Current observation techniques are able to probe the atmosphere of some giant exoplanets and get some clues about their atmospheric composition. However, the chemical compositions derived from observations are not fully understood, as for instance in the case of the \ce{CH4}/CO abundance ratio, which is often inferred different from what has been predicted by chemical models. Recently, the warm Neptune GJ~3470b has been discovered and because of its close distance from us and high transit depth, it is a very promising candidate for follow up characterisation of its atmosphere.}
{We study the atmospheric composition of GJ~3470b in order to compare with the current observations of this planet, to prepare the future ones, but also as a typical case study to understand the chemical composition of warm (sub-)Neptunes. The metallicity of such atmospheres is totally uncertain, and vary probably to values up to 100 $\times$ solar. We explore the space of unknown parameters to predict the range of possible atmospheric compositions.}
{We use a one-dimensional chemical code to compute a grid of models, with various thermal profiles, metallicities, eddy diffusion coefficient profiles, and stellar UV incident fluxes. Thanks to a radiative transfer code, we then compute the corresponding emission and transmission spectra of the planet and compare them with the observational data already published.}
{Within the parameter space explored we find that in most cases methane is the major carbon-bearing species. We however find that in some cases, typically for high metallicities with a sufficiently high temperature the \ce{CH4}/CO abundance ratio can become lower than unity, as suggested by some multiwavelength photometric observations of other warm (sub-)Neptunes, such as GJ~1214b and GJ~436b. As for the emission spectrum of GJ~3470b, brightness temperatures at infrared wavelengths may vary between 400 and 800 K depending on the thermal profile and metallicity.}
{Combined with a hot temperature profile, a substantial enrichment in heavy elements by a factor of $\geq$ 100 with respect to the solar composition can shift the carbon balance in favour of carbon monoxide at the expense of methane. Nevertheless, current observations of this planet do not allow yet to determine which model is more accurate.}


\maketitle

\section{Introduction}

In the recent past, multiwavelength observations of transiting exoplanets have been used to provide the first constraints on the chemical composition of exoplanet atmospheres. The identification of atmospheric constituents is currently restricted to gas giant planets with small orbital distances, because of the large transit depth variations. Most such efforts have concentrated on Jupiter size planets that orbit around solar-type stars, the so-called hot Jupiters. These planets are heavily irradiated by the nearby (early K-, G-, or late F-type) star, resulting in planetary equilibrium temperatures in excess of 1000~K. Transmission and dayside emission spectra of hot Jupiters such as HD~209458b and HD~189733b have revealed the presence of molecules such as CO, \ce{H2O}, \ce{CH4}, and \ce{CO2} in their atmospheres \citep{tin2007, gri2008, swa2008, swa2009b, swa2009a, mad2009}, although contradictory conclusions among different studies are not rare. Chemical models of hot Jupiter atmospheres which incorporate to a different degree processes such as thermochemical kinetics, vertical mixing, horizontal transport, and photochemistry \citep{lin2010, mos2011, kop2011, ven2012, agu2012} indicate that in such a hot hydrogen-helium dominated atmosphere, carbon monoxide and water vapour should be the major reservoirs of carbon and oxygen, while methane and carbon dioxide would be somewhat less abundant.

\begin{figure}
\centering
\includegraphics[angle=0,width=1.05\columnwidth]{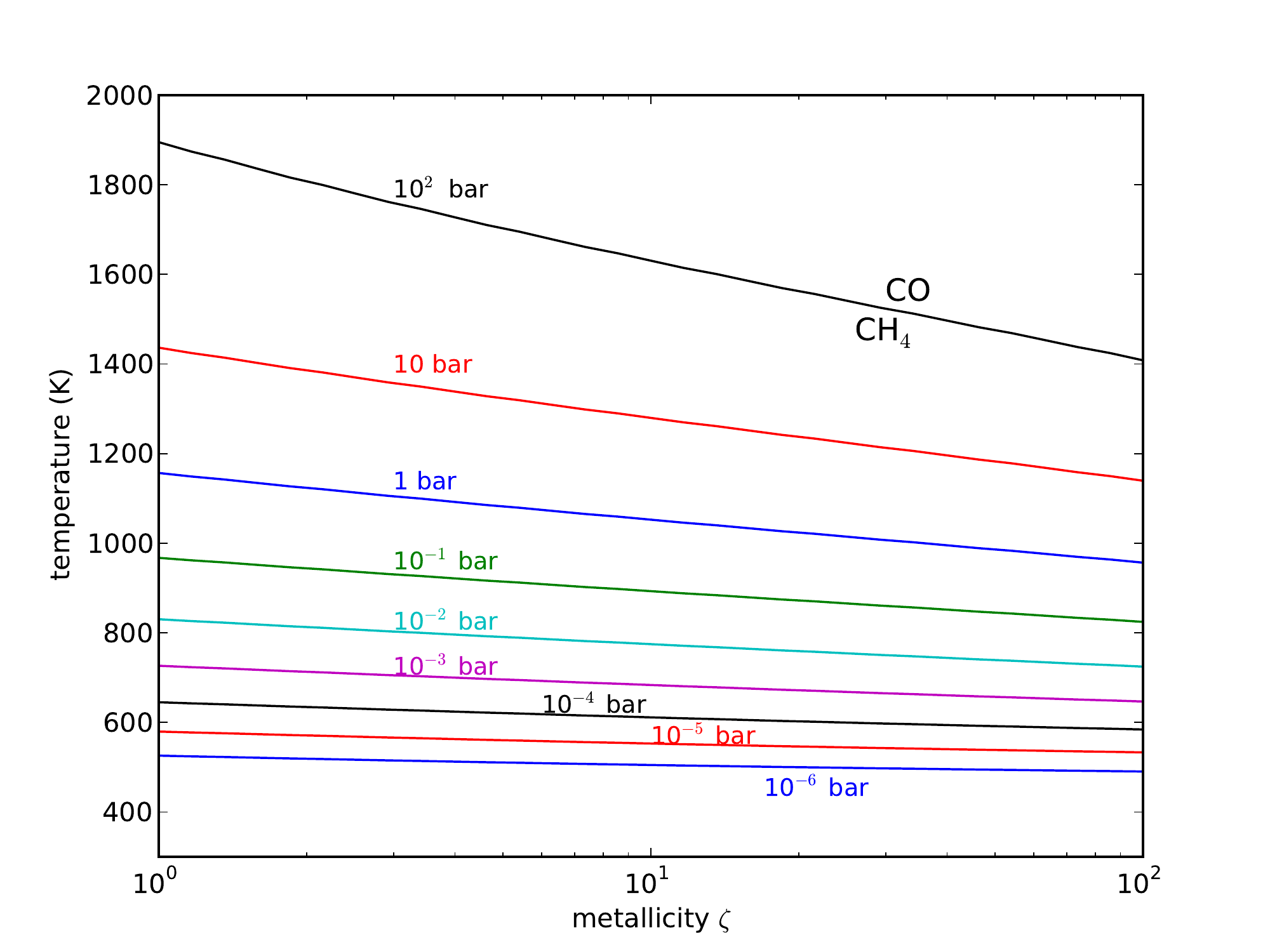}
\caption{Transition temperature for the C-bearing species (e.g. CO/\ce{CH4}) at the thermochemical equilibrium depending of metallicity, at different pressures. Above the curve, CO is the dominant C-bearing species, while \ce{CH4} dominates below. The temperature of the transition decreases when metallicity increases, an effect which is more important for high pressures.} \label{fig:COCH4}
\end{figure}

Even more challenging, transit spectra have recently allowed to characterise the atmosphere of the Neptune size planet GJ~436b and the mini Neptune or super-Earth GJ~1214b, both orbiting around M dwarf stars. Unlike Jupiter size planets, which have a very low occurrence rate around M dwarf stars \citep{joh2007, bon2013}, (sub-)Neptune size planets are found around both solar-type and M dwarf stars, although they are more easily observed around the latter type of stars because of the higher planet-to-star contrast which favour primary and secondary transit observations. These planets have at least a couple of interesting differences with respect to hot Jupiters. The first is related to the fact that the host M dwarf star is smaller and significantly cooler than a solar-type star, so that the planet is less severely heated (even if the orbital distances, in the range 0.01 -- 0.04 AU, are as small as for hot Jupiters), resulting in planetary effective temperatures below 1000~K. Interestingly, it is around this temperature that gaseous mixtures with solar elemental abundances show, under thermochemical equilibrium and at pressures around 1 bar, a sharp transition concerning the major carbon reservoir, CO and \ce{CH4} being the dominant carbon-containing species above and below 1000~K, respectively (Fig.~\ref{fig:COCH4}). In this regard it is interesting to note that transit spectra of GJ~436b indicates that its atmosphere is poor in methane \citep{ste2010, mad2011, knu2011}, yet this species is predicted to be the major carbon reservoir at thermochemical equilibrium. Such interpretation has been however disputed by \cite{bea2011} based on a different analysis of transmission spectra. 
A detailed chemical model by \cite{lin2011}, which considered thermochemical kinetics, vertical mixing, and photochemistry, concluded that \ce{CH4} should be the major carbon-bearing molecule in GJ~436b's atmosphere under most plausible conditions.

A second important difference with respect to hot Jupiters is that the lower mass of (sub-)Neptune planets allows to expect an elemental atmospheric composition significantly enriched in heavy elements, with respect to the solar composition, because of their lower efficiency to retain light elements \citep{elk2008}. In the case of GJ~1214b, its flat transmission spectrum indicates that the planet atmosphere either is hydrogen dominated but contains clouds or hazes, or has a high mean molecular weight, for instance an \ce{H2O}-rich atmosphere \citep{bean2010, bean2011, des2011, cro2011, crossfield2011, dem2012, ber2012}. The possibility of a hydrogen dominated atmosphere for GJ~1214b has been explored through chemical modelling by \cite{mil2012}, who found that methane would be the major carbon reservoir, just as the findings of \cite{lin2011}'s model on the atmosphere of GJ~436b, and that photolysis of \ce{CH4}, which could lead to the formation of hazes, would take place at heights substantially higher than required by the observations. These previous photochemical studies dedicated to (sub-)Neptunes explored high metallicities up to 50 $\times$ solar metallicity. In the solar system, Neptune and Uranus atmospheres have indeed carbon abundances about 50 times higher than in the Sun (oxygen being trapped in the deep and hot layers of the atmospheres which cannot be probed yet by observations). This carbon abundance is significantly higher than that of the atmosphere of Jupiter and Saturn, which is about 3 times solar \citep{her2004}. The bulk metallicity of icy giants is, however, much higher than that of their atmosphere, as they consist in a large fraction of rocks and ice \citep[e.g.][]{pollack1996, alibert2005}. In a warm Neptune that would have the same bulk composition as Neptune or Uranus, a larger fraction of the ices would be in the form of gases in the atmosphere, which may no longer be dominated by H$_2$ and He. Although the mass and radius of the planet, derived from radial velocities and transit measurements, can be used to constrain the bulk metallicity they do not provide a constraint on the metallicity of the envelope. This was shown by \citet{Baraffe2008} who modelled the evolution of Jupiter- and Neptune-mass planets with all the heavy elements in the core or distributed uniformly in the whole planet. They also used different approximations to model the equation of state of the enriched envelope. Their conclusion is that although planets with a uniform enrichment tend to have a smaller radius after about 1~Gyr compared with those with a core, the uncertainty related with the equation of state exceeds this difference. Therefore, the mass and radius of a planet may show that a large fraction of the planet mass consists in H$_2$-He but does not tell whether the envelope and the atmosphere are dominated by these compounds. The atmospheric abundance of H$_2$O, for instance, could reach or exceed that of H$_2$ giving the atmosphere a high mean molecular weight and small scale height. In this study, we considered an enrichment in heavy elements between 1 and 100 times solar. Using solar abundances from \citet{asp2009} and assuming that all the oxygen is in the form of H$_2$O, these enrichments correspond to a mean molecular mass between 2.3 and 4.1~g/mole. Heavy elements enrichment in the range 50-100 are extremely interesting because they correspond to a change of the carbon reservoir (either CH$_4$ or CO) for pressures within 1 and 100 bar and temperatures within 1000 and 2000~K (see Fig.~\ref{fig:COCH4}). The deep atmospheric layers where such conditions are found can contaminate most of the atmosphere due to the chemical quenching associated with vertical mixing \citep[e.g.][]{prinn1977carbon, lewis1984vertical, vis2011, mos2011, ven2012}.

In this work, we address the effects of the heavy elements enrichment in the transiting warm Neptune GJ~3470b discovered by \cite{bon2012}. This planet is a promising candidate for follow-up characterisation of its atmosphere and for a better understanding of the atmospheric chemistry of (sub-)Neptunes. GJ~3470b has a mass of 14~$M_{\oplus}$, in between those of GJ~436b (23~$M_\oplus$; \citealt{sou2010}) and GJ~1214b (6~$M_\oplus$; \citealt{har2013}). Its radius of 4.2~$R_\oplus$ implies a high amount of hydrogen in the envelope. Indeed, a planet with the same mass but made only of water would have half this radius. Some spectroscopic observations during primary transit have already been held during the past few months \citep{dem2013, fuk2013, crossfield2013, nasc2013}, leading sometimes to different and contradictory interpretations: hazy, cloud-free, metal-rich, low mean molecular weight, etc. Thus, more precise observations are needed to characterise its atmospheric structure and composition. While we wait for future observations, we study composition of the atmosphere of GJ~3470b with a model that includes  thermochemical and photochemical kinetics and vertical mixing. We explore the influence of the thermal profile, the vertical mixing efficiency, the poorly constrained UV irradiation and the metallicity on the chemical composition. We compute the resulting transmission and emission spectra that we compare with the observations available so far. While this work was being finalised, a similar study has been published on GJ~436b by \cite{mos2013}. We do not compare in details our results to theirs, but they are globally in agreement.

\section{The model}

\begin{table}
\caption{GJ~3470b's model parameters.} \label{table:star_planet_parameters}
\centering
\begin{tabular}{l@{\hspace{1.1cm}}r}
\hline \hline
Parameter                                                   & Value$^a$ \\
\hline
Stellar radius                                              & 0.503 $R_\odot$ \\
Stellar effective temperature                      & 3600~K \\
Planetary radius                                         & 4.2 $R_\oplus$ \\
Planetary mass                                          & 14.0~$M_\oplus$ \\
Planet-star distance                                  & 0.0348 AU \\
\hline
\end{tabular}
\tablenotea{$^a$ \cite{bon2012}.}
\end{table}

We aim at studying the atmospheric chemical composition in the dayside of GJ~3470b in the vertical direction. We have adopted the planetary and stellar parameters derived by \cite{bon2012}, which are given in Table~\ref{table:star_planet_parameters}. Note that the planetary parameters of GJ~3470b have been recently refined by \cite{dem2013} and \cite{fuk2013}, leading to a larger radius so to a smaller density than what was predicted first by \cite{bon2012}. These new observations imply that GJ~3470b has a quite low density ($\rho_p < 1$ g cm$^{-3}$) compared to Uranus and Neptune. The atmosphere model relies on some key input information such as the elemental composition, the vertical profile of temperature, the eddy diffusion coefficient, and the stellar ultraviolet (UV) flux, which are badly constrained. In order to explore to some extent the sensitivity of the atmospheric chemical composition to these uncertain parameters we have varied them around some standard choices. Hereafter we describe our choice of standard parameters and the range over which they have been varied.

\subsection{Stellar spectrum}\label{sec:stellar}
\begin{figure}
\centering
\includegraphics[angle=0,width=1.05\columnwidth]{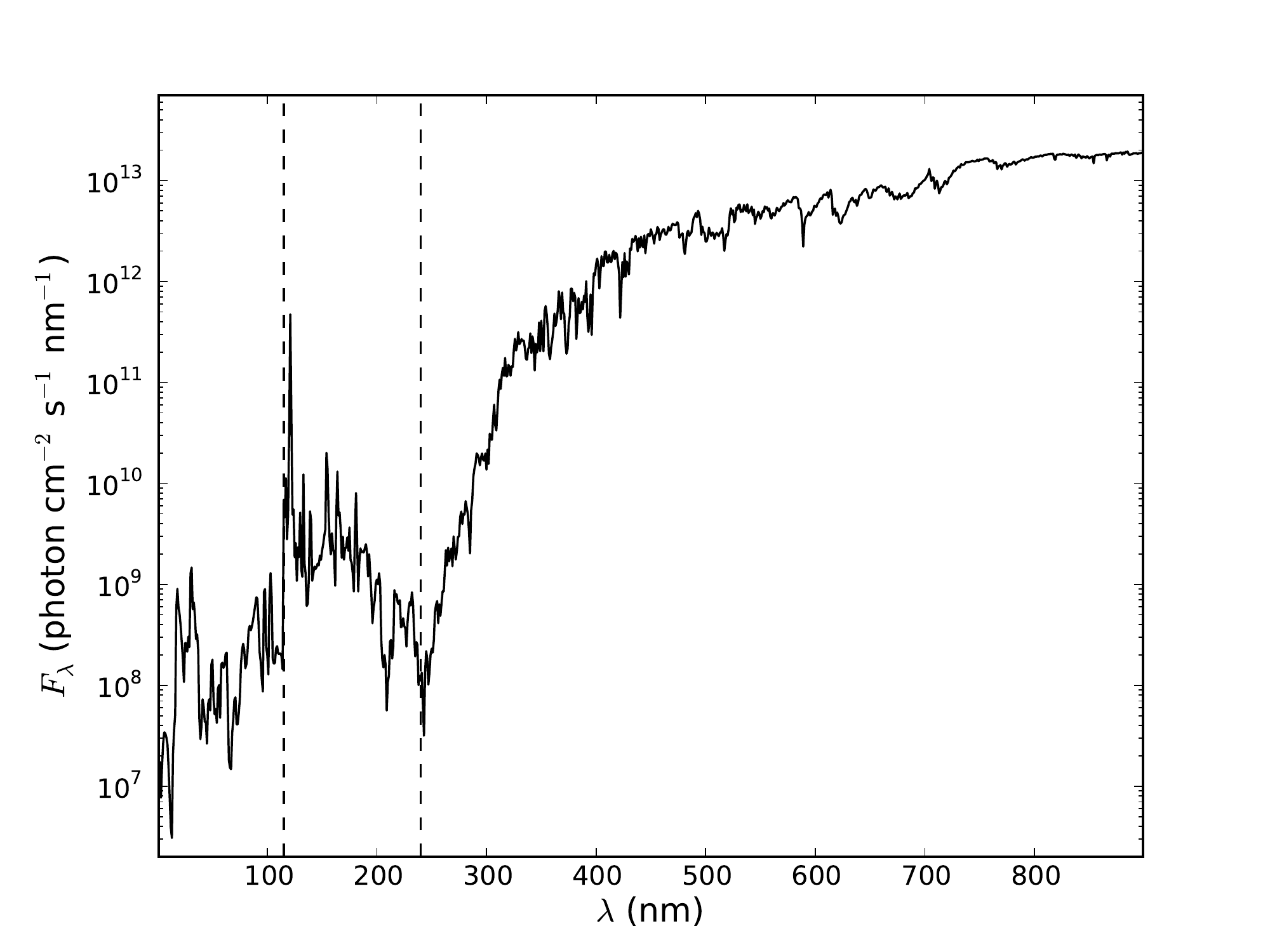}
\caption{Stellar spectrum adopted for GJ~3470, where the flux $F_\lambda$ is normalised to an orbital distance of 1~AU. Vertical dashed lines at 115 and 240~nm indicate the positions of junction between spectra from different sources (see text).} \label{fig:stellarspectrum}
\end{figure}

The radiation spectrum of the host star affects in two major ways the planetary atmosphere. On the one side, the visible-infrared part of the incoming stellar radiation controls the atmospheric thermal structure of the planet, and on the other, the UV radiation determines the photodissociation rates. Longward of 240~nm, we adopt a Phoenix NextGen synthetic spectrum \citep{hau1999} for a star with $T_{\rm eff}$~=~3600~K, $g$ = 10$^{4.5}$~cm s$^{-2}$, and solar metallicity. The flux of UV radiation emitted by M dwarf stars may vary by orders of magnitude depending of the degree of chromospheric activity of the star \citep{fra2013}. Unfortunately, the UV spectrum of GJ~3470 has not been observed to our knowledge, and we have therefore adopted the observed spectrum of the active M3.5V star GJ~644 \citep{seg2005} in the 115-240~nm wavelength range and the mean of Sun spectra at maximum and minimum activity \citep{gueymard2004} shortward of 115~nm. The final composite spectrum (shown in Fig.~\ref{fig:stellarspectrum}) is adopted as the standard stellar spectrum. Due to the large uncertainties at UV wavelengths, when exploring the space of parameters, we allow for a variation in the UV flux between 0.1 and 10 times the standard spectrum.

\subsection{Atmospheric metallicity}

To understand the history of GJ~3470b, it would be important to constrain its atmospheric metallicity, which is currently very uncertain. As it has been explained in the Introduction, the atmosphere of this planet can be enriched. The host star of GJ~3470b has a metallicity slightly above the solar value, [Fe/H]~=~+0.2, according to \cite{dem2013}. In our standard model, we consider that the abundances of heavy elements (other than H and He) are enriched in GJ~3470b's atmosphere by a factor of $\zeta =10$ with respect to the solar values compiled by \cite{asp2009}. However, given the large uncertainties in the elemental composition, when exploring the effect of metallicity on the atmospheric chemical composition we choose two extreme cases with $\zeta=1$ (solar metallicity) and $\zeta=100$ (high metallicity). In this study, we do not change the C/N/O relative abundance ratios compared to their solar values.

\subsection{Thermal profile}

\begin{figure}
\centering
\includegraphics[angle=0,width=1.05\columnwidth]{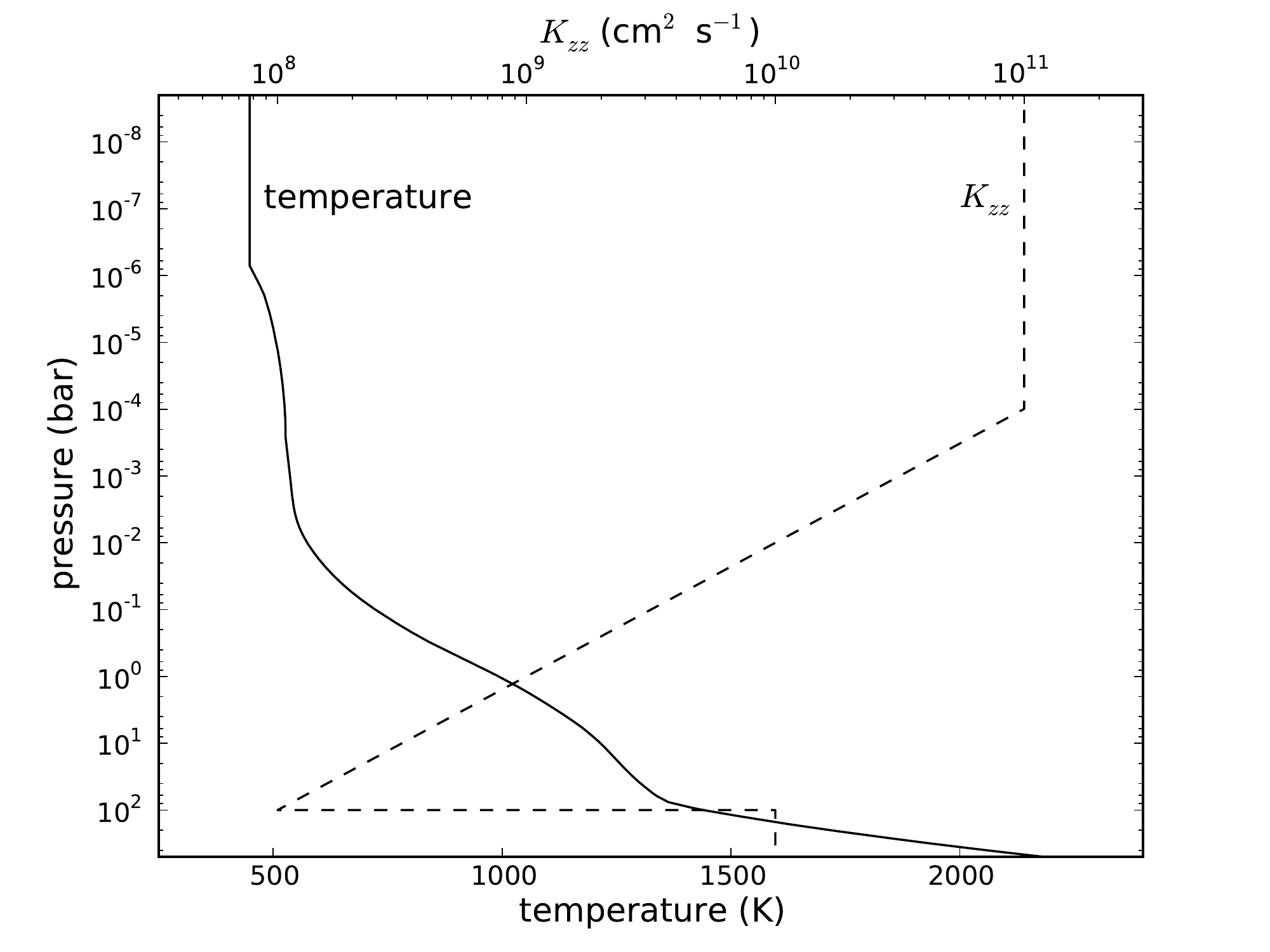}
\caption{Standard vertical profile of temperature (solid line referred to the lower abscissa axis) and of eddy diffusion coefficient (dashed line referred to the upper abscissa axis) adopted for the atmosphere of GJ~3470b.} \label{fig:profile_pTKzz}
\end{figure}

The vertical profile of temperature in GJ~3470b's atmosphere is computed with the radiative-convective model described by \cite{iro2005}, with the update of \cite{agu2012}. We adopt the planetary and stellar parameters given in Table~\ref{table:star_planet_parameters} as well as the input information corresponding to our standard model. The mixing ratios of the main species that provide opacity are estimated through thermochemical equilibrium, which is expected to be a good approximation as long as the abundances of CO and \ce{H2O} (the main species that affect the thermal structure) are close to the chemical equilibrium values. As seen in Section~\ref{subsec:results_standard}, this is likely to be the case for \ce{H2O} although not for CO throughout a good part of the atmosphere, which may add an uncertainty to the calculated thermal profile. In the deep atmosphere, the temperature is regulated by convective, rather than radiative, processes, and the internal flux of the planet becomes the most relevant parameter. The internal flux of the planet is highly uncertain since it depends on the age of the planet and on processes of dissipation of energy which may be triggered by for instance tidal effects \citep{agu2013}. We have adopted an internal flux which corresponds to an internal temperature of 100~K, a value commonly used in previous studies in the absence of relevant constraints. The temperature is calculated vertically as a function of pressure between 1000 and 10$^{-6}$ bar, and above this latter pressure level an isothermal atmosphere is assumed. The calculated vertical profile of temperature, which is adopted as the standard one, is shown in Fig.~\ref{fig:profile_pTKzz}. Given the various uncertainties that affect the calculated temperature profile, we explore it in our space of parameters choosing two bounding cases in which a value of 100~K is added and subtracted to the standard temperature profile.

\subsection{Vertical mixing}
Another important parameter for the chemical model is the vertical profile of the eddy diffusion coefficient, which determines the efficiency of the vertical mixing as a function of pressure. In the case of exoplanet atmospheres, constraints on this parameter come solely from global circulation models (GCMs). For the atmosphere of GJ~3470b we adopt a parametric profile for the eddy diffusion coefficient, with a high value of $K_{zz}=10^{10}$~cm$^2$s$^{-1}$ in the convective region of the atmosphere (which is approximately located below the 100 bar pressure level), and values inferred from the GCM of GJ~436b developed by \cite{lew2010}. By multiplying a mean vertical wind speed by the local scale height, these authors estimated $K_{zz}$ values of 10$^8$~cm$^2$s$^{-1}$ at 100~bar and 10$^{11}$~cm$^2$s$^{-1}$ at 0.1~mbar. We have therefore adopted these values and assumed a linear behaviour in the logarithm of $K_{zz}$ with respect to the logarithm of pressure in the 10$^{-4}$ -- 100~bar regime, and a constant value for $K_{zz}$ at higher atmospheric layers. The resulting vertical profile, which we adopt as the standard one, is shown in Fig.~\ref{fig:profile_pTKzz} referred to the upper abscissa axis. However, because the GCM of \cite{lew2010} is constructed for GJ~436b and not for GJ~3470b, and also because the method used to estimate the eddy diffusion coefficient is highly uncertain \citep[e.g.][]{par2013}. We have explored the sensitivity of the chemical abundances to the eddy diffusion coefficient and consider two limiting cases in which $K_{zz}$ is divided and multiplied by a factor of ten with respect to the standard profile above the convective region.

\subsection{Kinetics}
Once the physical parameters and elemental composition are established, the atmospheric chemical composition is computed by solving the equation of continuity in the vertical direction for 105 species composed of H, He, C, N, and O. The reaction network and photodissociation cross sections used are described in \cite{ven2012}. This chemical network, which includes $\sim$1000 reversible reactions (so a total of $\sim$2000 reactions), has been developed from applied combustion models and has been validated over a large of temperature (from 300 to 2500~K) and pressure (from a few mbar to some hundred of bar). It is able to reproduce the kinetic evolution of species with up to 2 carbon atoms. Thus, our chemical network is valid to study the chemical composition of the atmosphere of GJ~3470b.

We can compare our results with previous results obtained for other (sub-)Neptune atmospheres, as for instance \cite{lin2011} (GJ~436b) and \cite{mil2012} (GJ~1214b). Both studies use smaller chemical networks than ours ($\sim$700 reactions and 51 and 61 species for respectively \citealt{lin2011} and \citealt{mil2012}) and reverse all reaction rates using the principle of microscopic reversibility \citep{vis2011, ven2012}. However, contrary to our network, none of them have been validated as a whole through experiments. \cite{lin2011} use the chemical network conceived for Jovian planets \citep[and reference therein]{liang2003, liang2004} updated for high temperature \citep{lin2010}, enhanced with nitrogen reactions and a small set of \ce{H2S} reactions. \cite{mil2012} use chemical network of \cite{zahnle2009soots}, so also originally made for Jovian planet \citep{zahnle1995} and upgraded for high temperature atmospheres with an arbitrary selection of new reaction rates from available data \citep{zahnle2009atmospheric}. As it has been shown in \cite{ven2012}, different chemical schemes can lead to different quenching levels and thus to differences in computed atmospheric composition. Thus, some differences found between, on one hand, this study and, on the other hand, \cite{lin2011} and \cite{mil2012}, may be due to the use of different chemical schemes.

\begin{table}
\caption{Model's parameter space explored. All the parameters are changed with respect to the standard values showed in Figs.~\ref{fig:stellarspectrum} and \ref{fig:profile_pTKzz}. The standard metallicity is 10 $\times$ solar ($\zeta=10$).} \label{table:models_grid}
\centering
\begin{tabular}{llll}
\hline \hline
Parameter      & Range of values                                         & Symbol \\
\hline
Metallicity     & Solar ($\zeta=1$)       & $\zeta_1$ \\
                     & High ($\zeta = 100$) & $\zeta_{100}$ \\
\hline
Temperature & Warm atmosphere ($+100$~K)        & $T_{+100}$ \\
                     & Cool atmosphere ($-$100~K)       & $T_{-100}$ \\
\hline
Eddy diffusion coefficient & High ($K_{zz}$ $\times 10$)     & $K_{zz}^{\times 10}$ \\
                                         & Low ($K_{zz}$ $\div 10$)           & $K_{zz}^{\div 10}$ \\
\hline
Stellar UV flux                   & High irradiation ($F_{\lambda}$ $\times 10$) & $F_{\lambda}^{\times 10}$ \\
                                         & Low irradiation ($F_{\lambda}$ $\div 10$)       & $F_{\lambda}^{\div 10}$ \\
\hline
\end{tabular}
\end{table}

\section{Results and discussion} \label{sec:results}

Our standard set of parameters to build up the chemical model of GJ~3470b's atmosphere consists of an elemental composition given by $\zeta = 10$, the vertical profiles of temperature and eddy diffusion coefficient shown in Fig.~\ref{fig:profile_pTKzz}, and the stellar UV spectrum shown in Fig.~\ref{fig:stellarspectrum}. Apart from this standard model we have constructed a grid of 16 models in which we have explored the sensitivity of the chemical composition to the metallicity, temperature, eddy diffusion coefficient, and stellar UV flux, according to the choices detailed in Table~\ref{table:models_grid}. 
For all of the seventeen models, the initial conditions are the thermochemical equilibrium. At both upper and lower boundaries, we impose a zero flux for each species. The steady-state is reached after an integration time of $t=10^8$s ($K_{zz}^{\times 10}$) or $t=10^9$s ($K_{zz}^{\div 10}$).

\subsection{Standard model} \label{subsec:results_standard}

\begin{figure}
\centering
\includegraphics[angle=0,width=1.05\columnwidth]{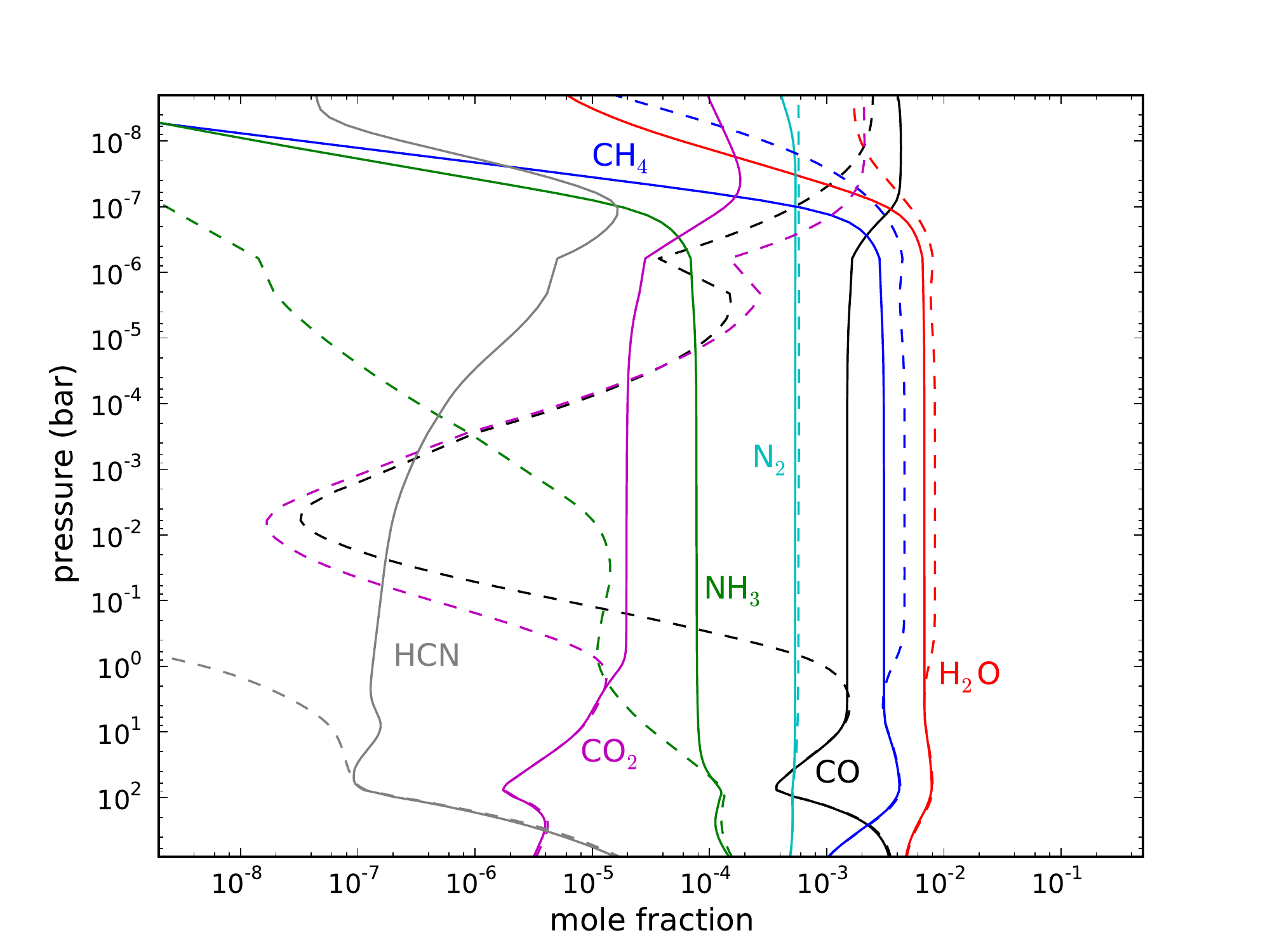}
\caption{Vertical distribution of molecular abundances in the standard model of GJ~3470b's atmosphere as computed through thermochemical equilibrium (dashed lines) and with the model that includes thermochemical kinetics, vertical mixing, and photochemistry (solid lines).} \label{fig:abundances_standard}
\end{figure}

In this section, we present the results of our standard model and compare them with previous publications dealing with (sub-)Neptunes: \cite{lin2011} on GJ~436b and \cite{mil2012} on GJ~1214b. Because these models do not use the same thermal profiles as us, nor the same eddy diffusion profiles and elemental abundances, it is difficult to compare quantitatively our results. Nevertheless, different cases have been studied in these publications so we can compare qualitatively the results that we obtained.

\subsubsection{Chemical composition}

Figure~\ref{fig:abundances_standard} shows the atmospheric composition of GJ~3470b at the chemical equilibrium (dashed lines) and at the steady-state, computed with the model taking into account thermochemical kinetics, vertical mixing, and photochemistry (solid lines). The abundances of all species remain at chemical equilibrium for pressures higher than about 40 bar, while at lower pressures we can see the effect of vertical mixing. Around 40 bar the abundances of HCN and \ce{NH3} depart from chemical equilibrium, and at somewhat lower pressure, around 2 bar, the abundances of \ce{CO2}, CO, \ce{CH4}, and \ce{H2O} get quenched, i.e. they are frozen at the chemical equilibrium value of the quench level. This quenching effect makes \ce{CH4}, \ce{H2O}, and \ce{N2} to be slightly less abundant than what thermochemical equilibrium predicts, so that CO, \ce{NH3}, \ce{CO2}, and HCN can be more abundant than the equilibrium prediction. In the upper atmosphere (above the $10^{-6}$ bar level), we see the effect of photodissociations: some species (for example \ce{H2O} and \ce{CH4}) are destroyed by photolysis, whereas other (as \ce{CO2} and \ce{CO}) see their abundance increased. Globally, between $10^2$ and $10^{-6}$ bar, the most abundant species of the atmosphere of GJ~3470b (after \ce{H2} and He) are, by decreasing order, \ce{H2O}, \ce{CH4}, and CO.

First, we compare our results with those of \cite{lin2011}. We focus on their cases where elemental abundances are solar and $50~\times$ solar. Our $T-P$ profile is not very different from their so we expect to have results quite similar. Even if our eddy diffusion coefficient is not identical, the abundances we find for all species are in between these two cases. In the region where vertical quenching dominates (in between the thermochemical equilibrium and photochemical regions) the behaviour of abundances is rather similar since the eddy diffusion coefficient adopted for the quenching level is not very different ($10^8$ cm$^2$ s$^{-1}$ by \cite{lin2011} and somewhat higher in our case). However, in the upper layers our adopted $K_{zz}$ value is substantially higher than the value of $10^8$ cm$^2$ s$^{-1}$ adopted by \cite{lin2011}, so that in their model the region where photochemistry takes place is shifted to lower heights.

Then, we compare our results with those obtained by \cite{mil2012} using $5~\times$ and $30~\times$ solar elemental abundances and an eddy diffusion coefficient of $K_{zz}=10^9$cm$^2$s$^{-1}$. We expect our results to be in between these two results. That is what we find for most species, except CO and \ce{CO2}. For these two species, at the steady-state, our model gives abundances about 100 times higher than in their case $\zeta = 30$. This is due to the fact that the abundances of these species depart from chemical equilibrium at a higher pressure in the study of  \cite{mil2012} than in ours ($\sim$ $10^{2}$ bar and $\sim$ 5 bar, respectively). Indeed, \cite{mil2012} use a thermal profile quite similar to ours, except for pressures higher than 1 bar. While in our $T-P$ profile the temperature increases with pressure, in theirs, the temperature remains constant between 1 and 100 bar. Consequently, the temperature in the deeper part of the atmosphere, where quenching happens, is colder than in our $T-P$ profile. This difference has consequences on the abundances of some species at the chemical equilibrium (for a given pressure level, CO and \ce{CO2} have equilibrium abundances smaller than in our model) and also at the steady-state because quenching happens at different levels.

\subsubsection{\ce{CH4}/CO abundance ratio}

The \ce{CH4}/CO abundance ratio is an important parameter to discuss, since some observational and modelling studies seem to indicate a poor methane content in the atmosphere of warm (sub-)Neptunes while thermochemical equilibrium predicts that \ce{CH4} should be the major carbon reservoir in such atmospheres (e.g \citealt{ste2010, mad2011, knu2011} for GJ~436b and \citealt{mil2012} for GJ~1214b). Of course chemical equilibrium depends on the $T-P$ profile and the assumed elemental composition, but this findings have suggested the need to invoke non-equilibrium processes such as mixing and photodissociations to help explaining these non expected chemical compositions. Nevertheless, even taking into account these non-equilibrium processes, 1D chemical models have not been able to find the set of parameters that may lead to a \ce{CH4}/CO abundance ratio lower than 1. In the case of the warm Neptune GJ~436b, observations of the dayside emission seem to indicate that this planet has an atmosphere dominated by CO and poor in \ce{CH4} (\ce{CH4}/CO abundance ratio  equals to $10^{-4}$ -- $10^{-3}$ for \citealt{ste2010} and \citealt{mad2011}), although a different interpretation has been provided by \cite{bea2011} based on a re-analysis of the same secondary eclipse data and on primary transit observations, which indicates a high methane content in the atmosphere, with eventually traces of CO or \ce{CO2}. Whatever the right interpretation, the chemical modelling done by \cite{lin2011} shows that \ce{CH4} is more abundant than CO (between $10^{-3}$ and 1 bar the \ce{CH4}/CO abundance ratio is $\sim2\times10^3$ and $\sim3$ for metallicities of $\zeta = 1$ and 50, respectively). In the case of GJ~1214b, \cite{mil2012} find that \ce{CH4} should be the major carbon reservoir in the atmosphere, with \ce{CH4}/CO abundance ratios of $\sim6\times10^4$ and $\sim10^3$ between $10^{-2}$ and 1 bar for metallicities of $\zeta = 5$ and 30, but find also that it is a no-\ce{CH4} model that best fits the observations of this planet.

With the standard value of the parameters of GJ~3470b, we find a \ce{CH4}/CO abundance ratio of 2 at 1 bar, i.e. with \ce{CH4} being slightly more abundant than CO. We then explore how the \ce{CH4}/CO abundance ratio varies within the space of parameters.

\subsection{The parameter space of $\zeta$, $T$, $K_{zz}$, and $F_\lambda$} \label{subsec:results_grid}

\begin{figure*}[!ht]
\centering
\includegraphics[angle=0,width=0.96\textwidth]{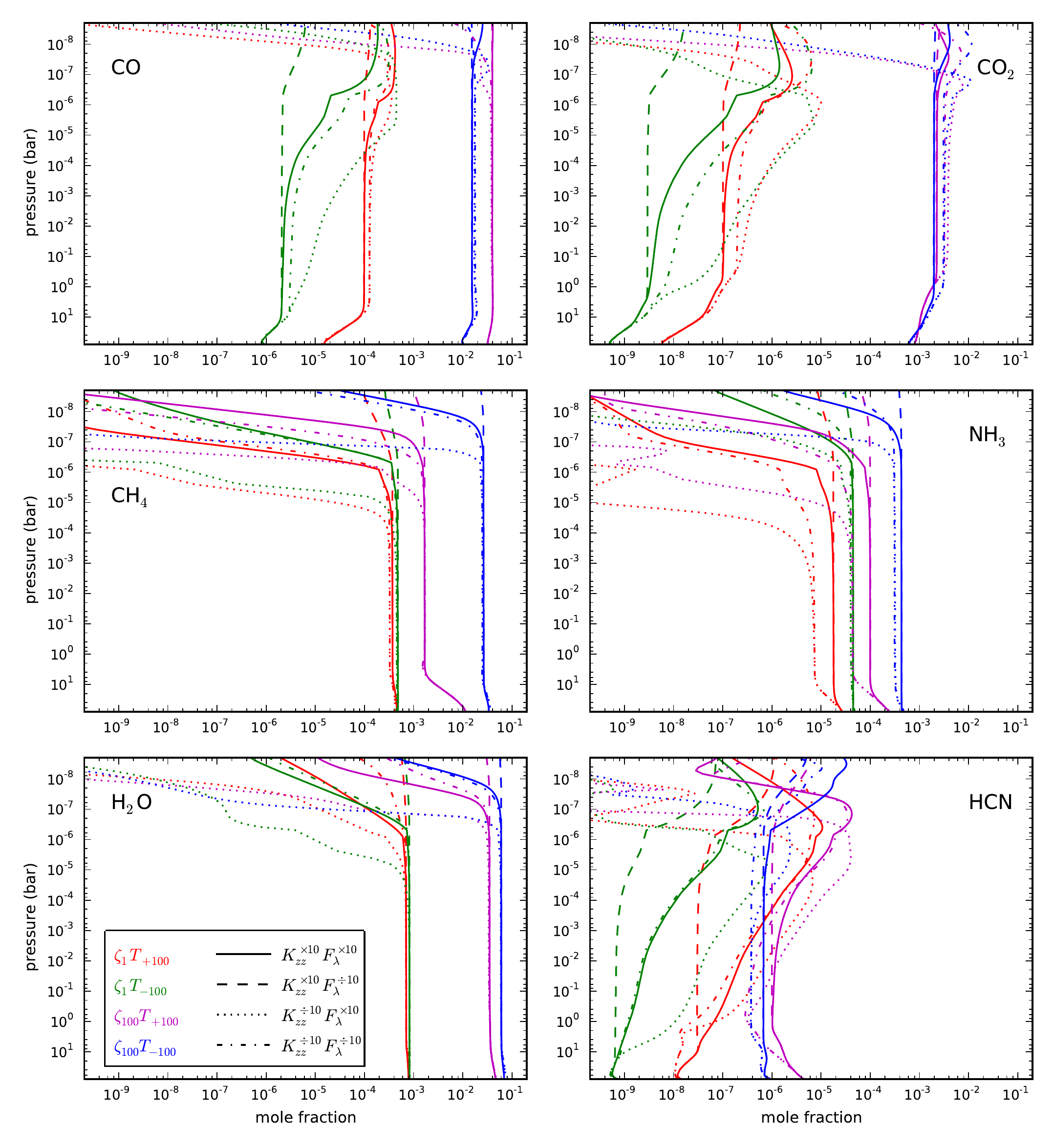}
\caption{Vertical distribution of the abundances of selected molecules as calculated through each of the 16 models in which the space of parameters of metallicity, temperature, eddy diffusion coefficient, and stellar UV flux are explored. Each colour corresponds to a set of metallicity and temperature, and each line style to a set of eddy diffusion coefficient and stellar UV flux (see legend in the \ce{H2O} panel and meaning of each symbol in Table~\ref{table:models_grid}).} \label{fig:abundances_16models}
\end{figure*}

We study the different possible atmospheric compositions of GJ~3470b by exploring the space of unknown parameters: metallicity ($\zeta$), temperature ($T$), eddy diffusion coefficient ($K_{zz}$), and incident UV flux ($F_\lambda$). The computed abundances of CO, \ce{CO2}, \ce{CH4}, \ce{NH3}, \ce{H2O} and HCN are plotted on Fig.~\ref{fig:abundances_16models}. We choose these species because they are the ones that most influence the planetary spectra.

\subsubsection{Effect of metallicity}

An increase in the metallicity obviously produces an abundance enhancement of all molecules containing heavy atoms (compare red and magenta lines, or green and blue lines, in Fig.~\ref{fig:abundances_16models}). Apart from this, depending of the molecule, the reaction to a change in the metallicity can be quite different, with the most sensitive one being carbon dioxide. When the metallicity changes from $\zeta=1$ to 100, the abundance of \ce{CO2} increases by a 4--6 orders of magnitude, while that of CO increases by 2--4 orders of magnitude, and the rest molecules experience less dramatic variations. A large abundance of \ce{CO2} would probably be the best evidence of an enhanced metallicity in the planet's atmosphere, as already found by \cite{zahnle2009soots} in the case of hot Jupiters. Nitrogen species are also sensitive to metallicity. \ce{N2} and HCN increase their abundance by $\sim$2 orders of magnitude when metallicity increases.

\subsubsection{Effect of temperature}

At the typical temperatures expected in the atmosphere of a warm Neptune such as GJ~3470b, a variation of temperature of 200 K can produce important changes in the resulting chemical abundances. These abundance variations depend to a large extent on the adopted metallicity. If we focus on the most abundant molecules, concretely on the six shown in Fig.~\ref{fig:abundances_16models}, there are two clear behaviours. On the one hand, we have CO, \ce{CO2}, and HCN, which experience an abundance enhancement when the temperature is increased, especially at low metallicities ($\zeta = 1$), in which case the abundances of these molecules vary by 1--2 orders of magnitude. We notice that at high metallicities ($\zeta = 100$), the change of temperature has less effect, the molar fraction of CO and HCN varying only by a factor $\sim$2 whereas the abundance of \ce{CO2} exhibits negligible change.
On the other hand, \ce{CH4}, \ce{H2O}, and \ce{NH3} respond to an increase of temperature in the opposite direction, i.e. decreasing their abundances. In this case the effect is more apparent at high metallicities, with abundance variations up to one order of magnitude.

The chemical composition in the atmosphere of warm (sub-)Neptunes can be quite sensitive to the temperature, especially if the temperatures in the quench region, usually located in the 0.1--10~bar pressure range, are around 1000~K, since at these temperatures and pressures there are important transitions such as that concerning CO and \ce{CH4}. Uncertainties in the thermal profile are therefore a major source of error in some of the calculated abundance ratios.

\subsubsection{Effect of vertical mixing}

The stronger vertical mixing is, the deeper quenching happens and the more the upper atmosphere will be contaminated by the chemical composition of the deep atmosphere. This has a crucial importance to interpret the observations and thus the composition of such atmospheres. Globally, a higher eddy diffusion coefficient results in a stronger vertical mixing, so in abundance profiles more flat in the vertical direction. 
For CO, \ce{CO2}, and HCN, at pressures above the quenching level, a high $K_{zz}$ leads to smaller abundances than with a low $K_{zz}$, whereas for all the other species, a high $K_{zz}$ creates globally higher abundances.

\subsubsection{Effect of stellar UV flux}

For some species (HCN, \ce{CO2} and CO), thanks to vertical mixing, the effect of photochemistry propagates quite deep in the atmosphere, down to few bars. For these species (if \ce{CH4} is reservoir of carbon), the effect of photochemistry is to enhance their abundance, especially at low metallicity. A more intense UV flux results in an increase of their abundance.
For the other species represented on Fig.~\ref{fig:abundances_16models}, the effect of the UV flux remains only at low pressures ($< 10^{-3}$ bar). These species are destroyed by photolysis, so a higher photochemistry shifts photodestruction of molecules to lower heights.

\begin{figure*}[!htbp]
\centering
\includegraphics[angle=0,width=0.45\textwidth]{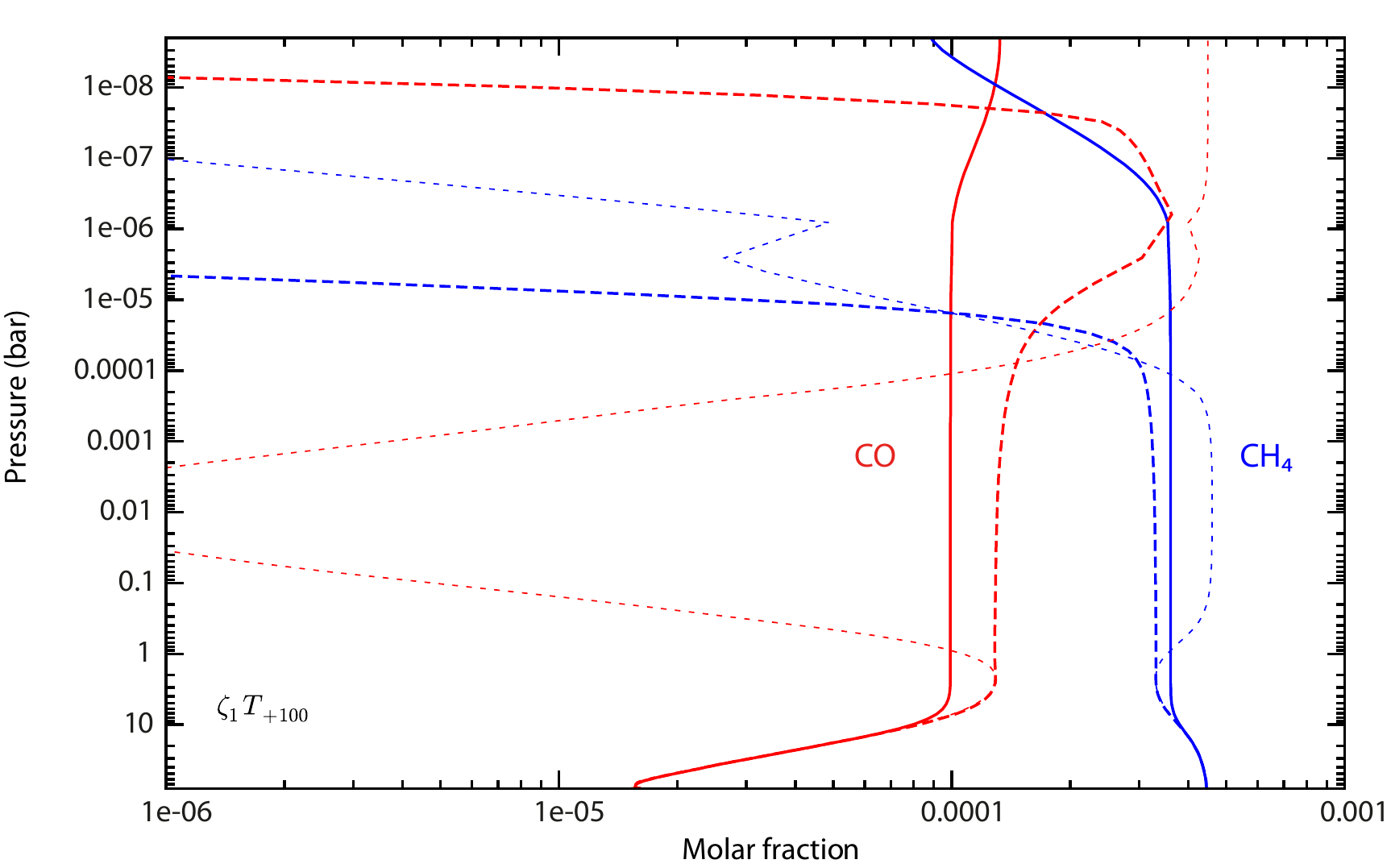}
\includegraphics[angle=0,width=0.40\textwidth]{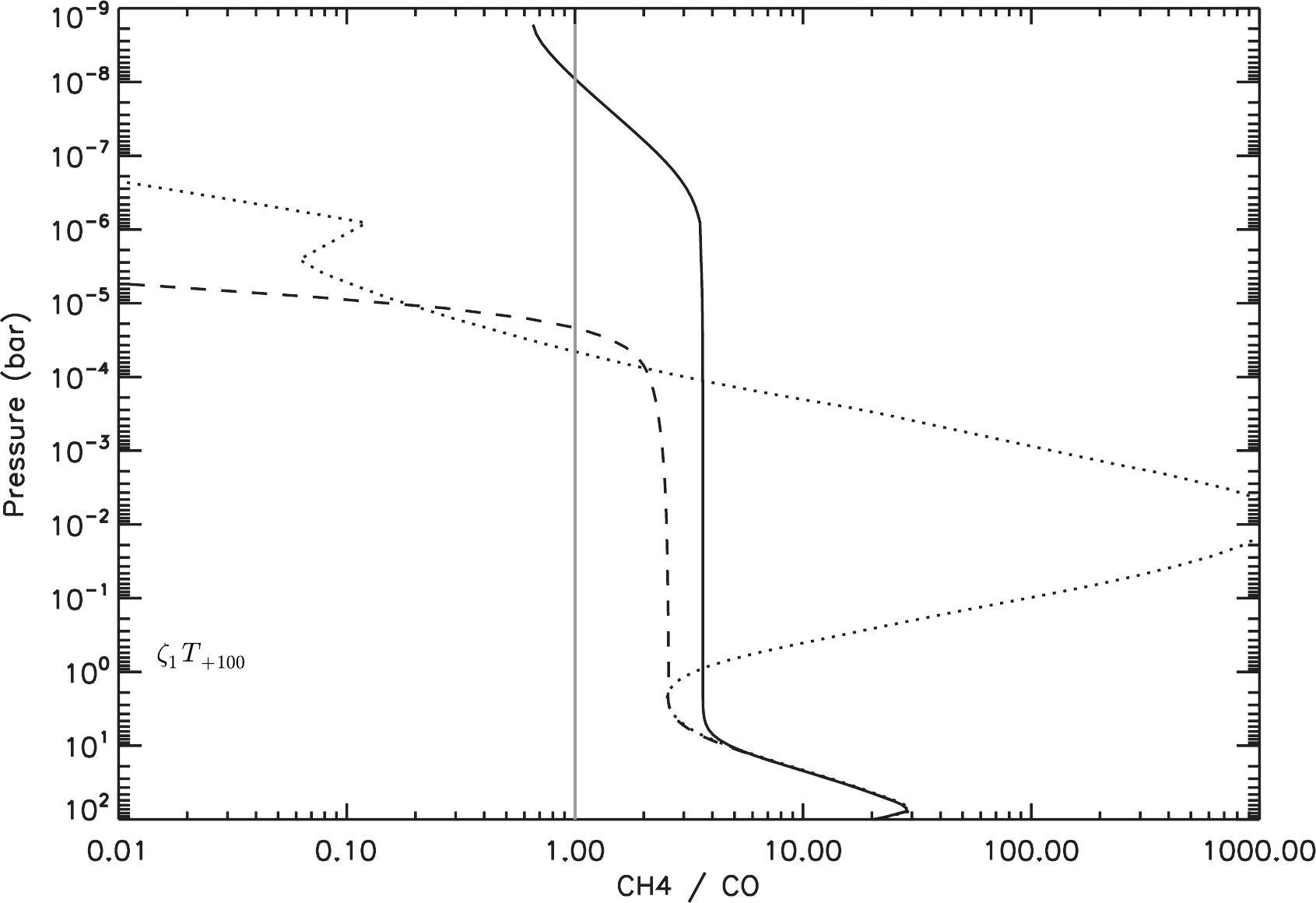}
\includegraphics[angle=0,width=0.45\textwidth]{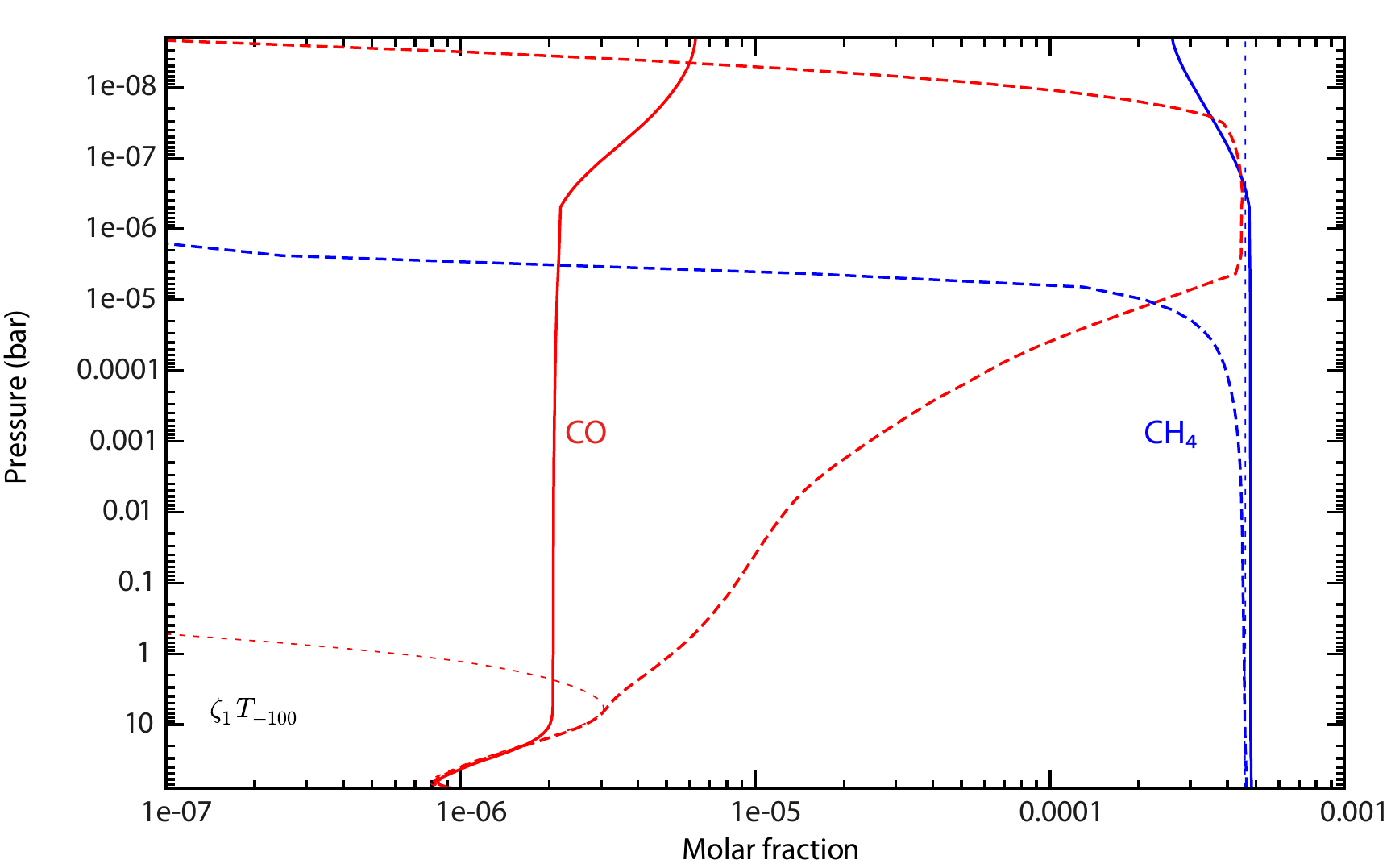}
\includegraphics[angle=0,width=0.40\textwidth]{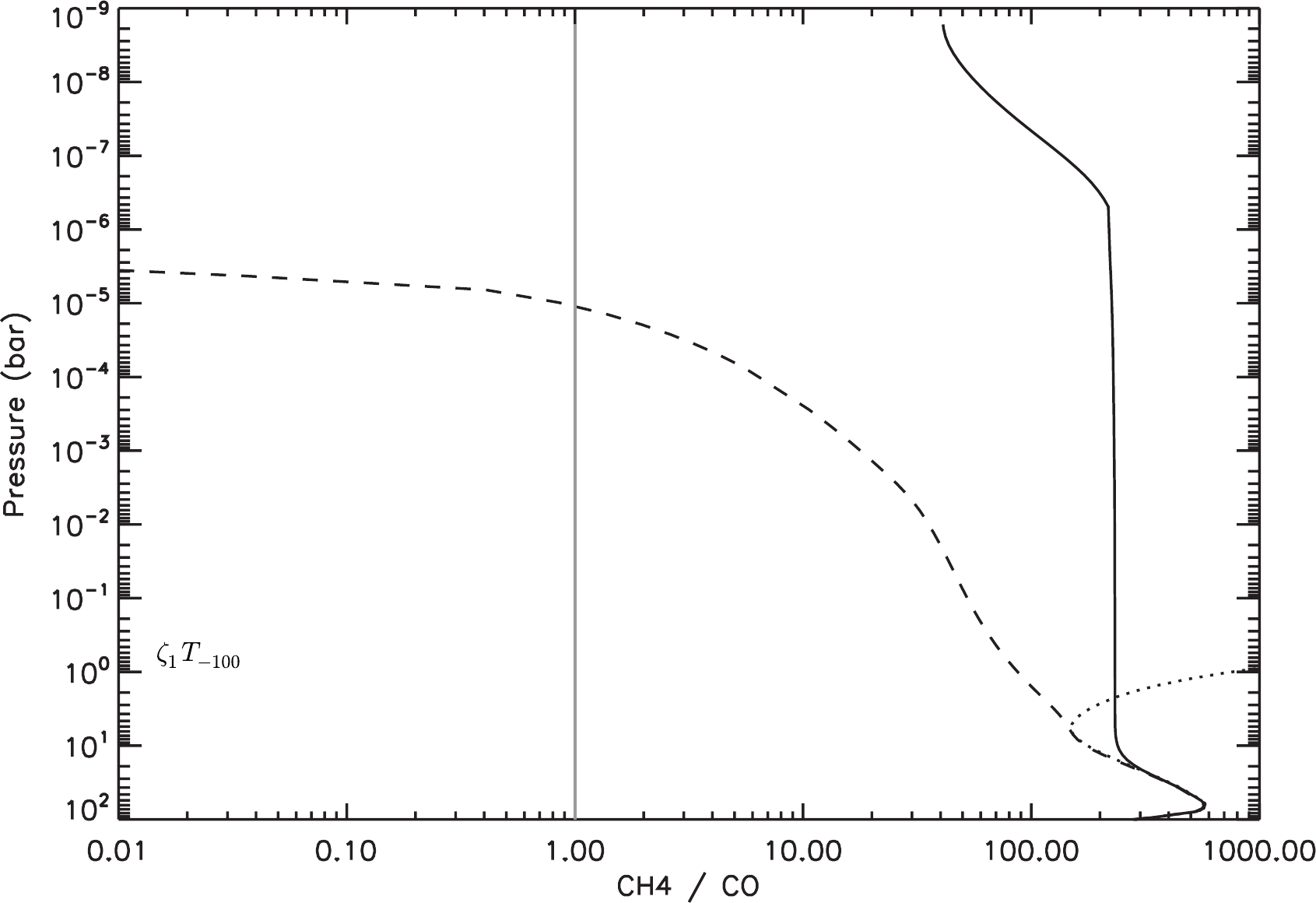}
\includegraphics[angle=0,width=0.45\textwidth]{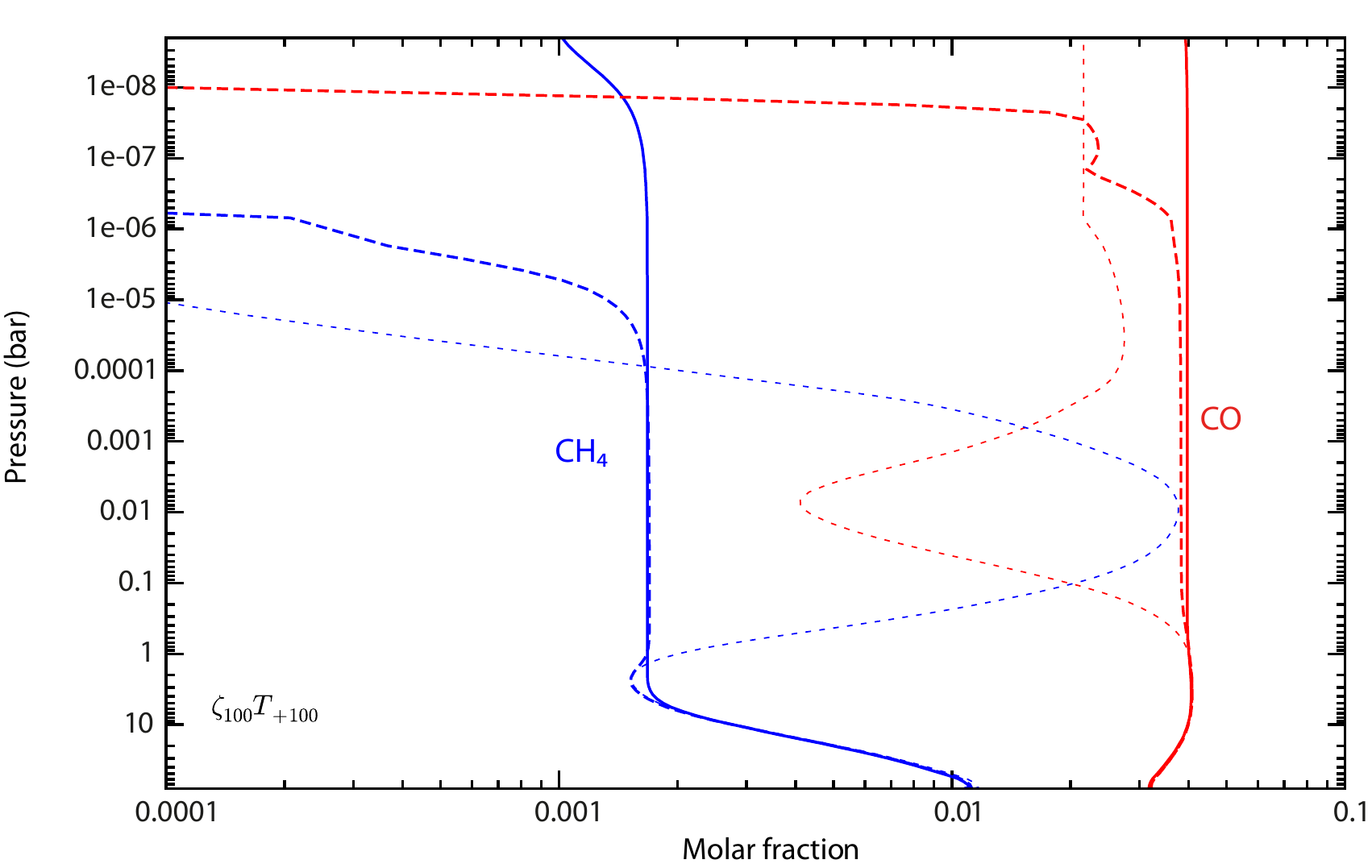}
\includegraphics[angle=0,width=0.40\textwidth]{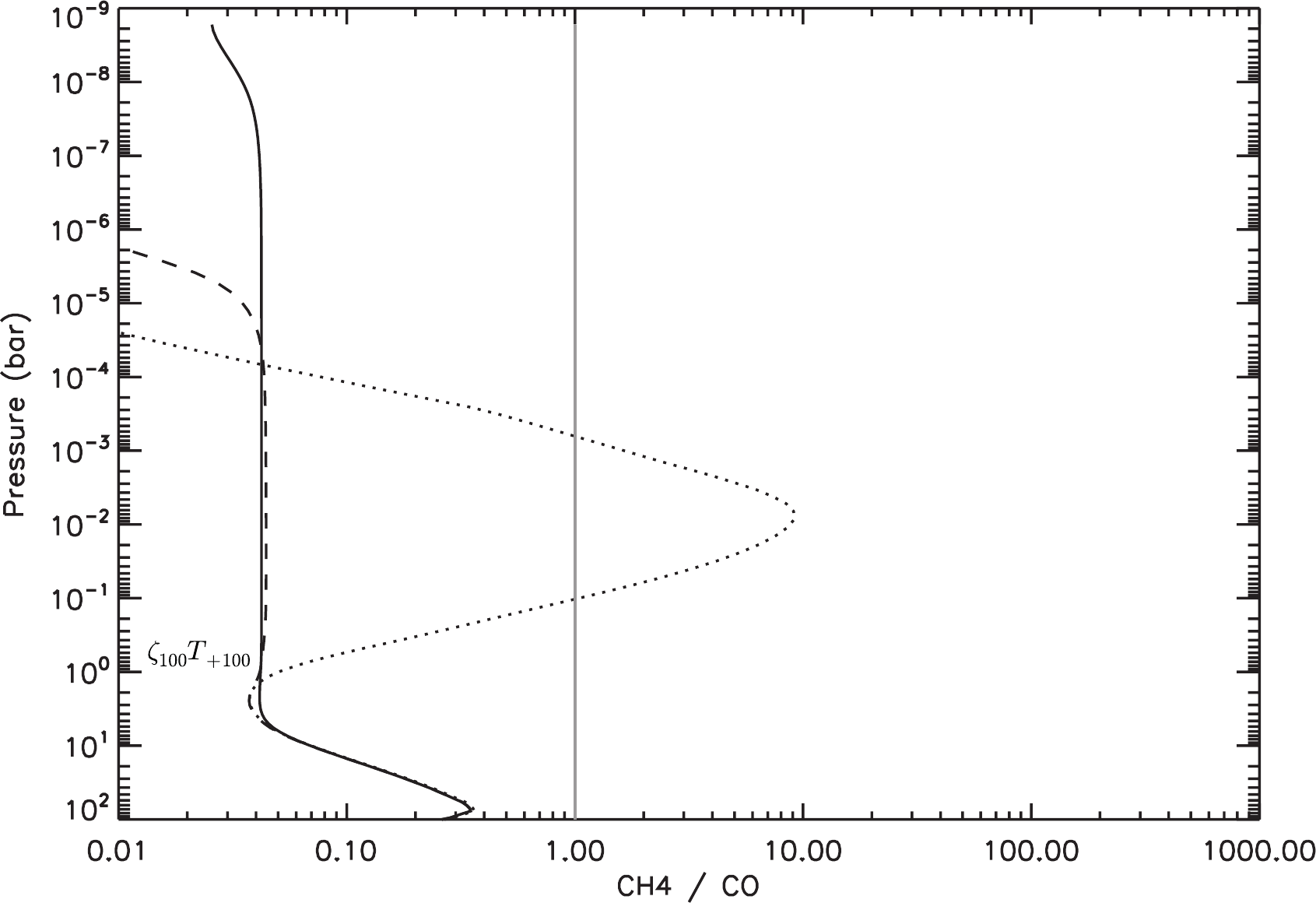}
\includegraphics[angle=0,width=0.45\textwidth]{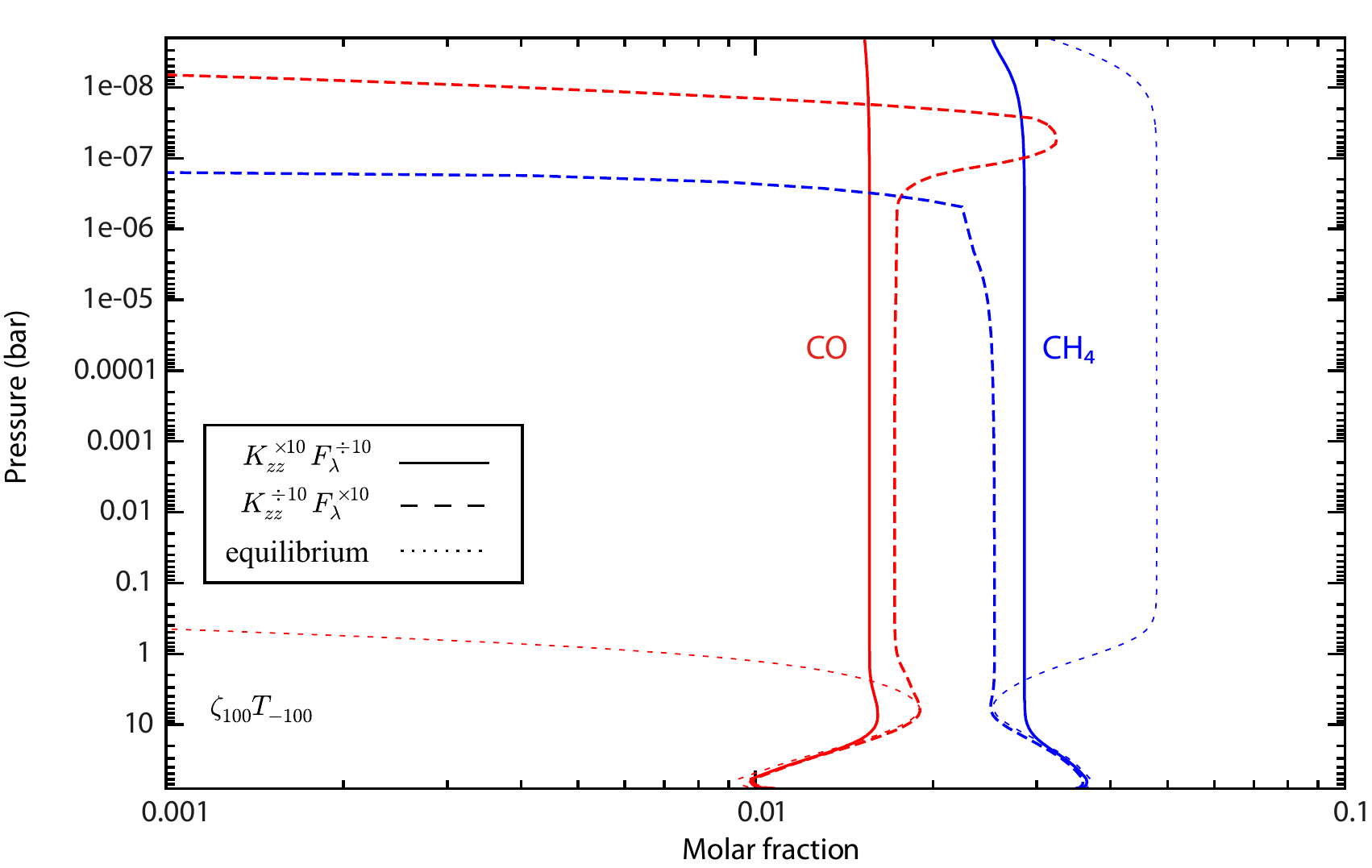}
\includegraphics[angle=0,width=0.40\textwidth]{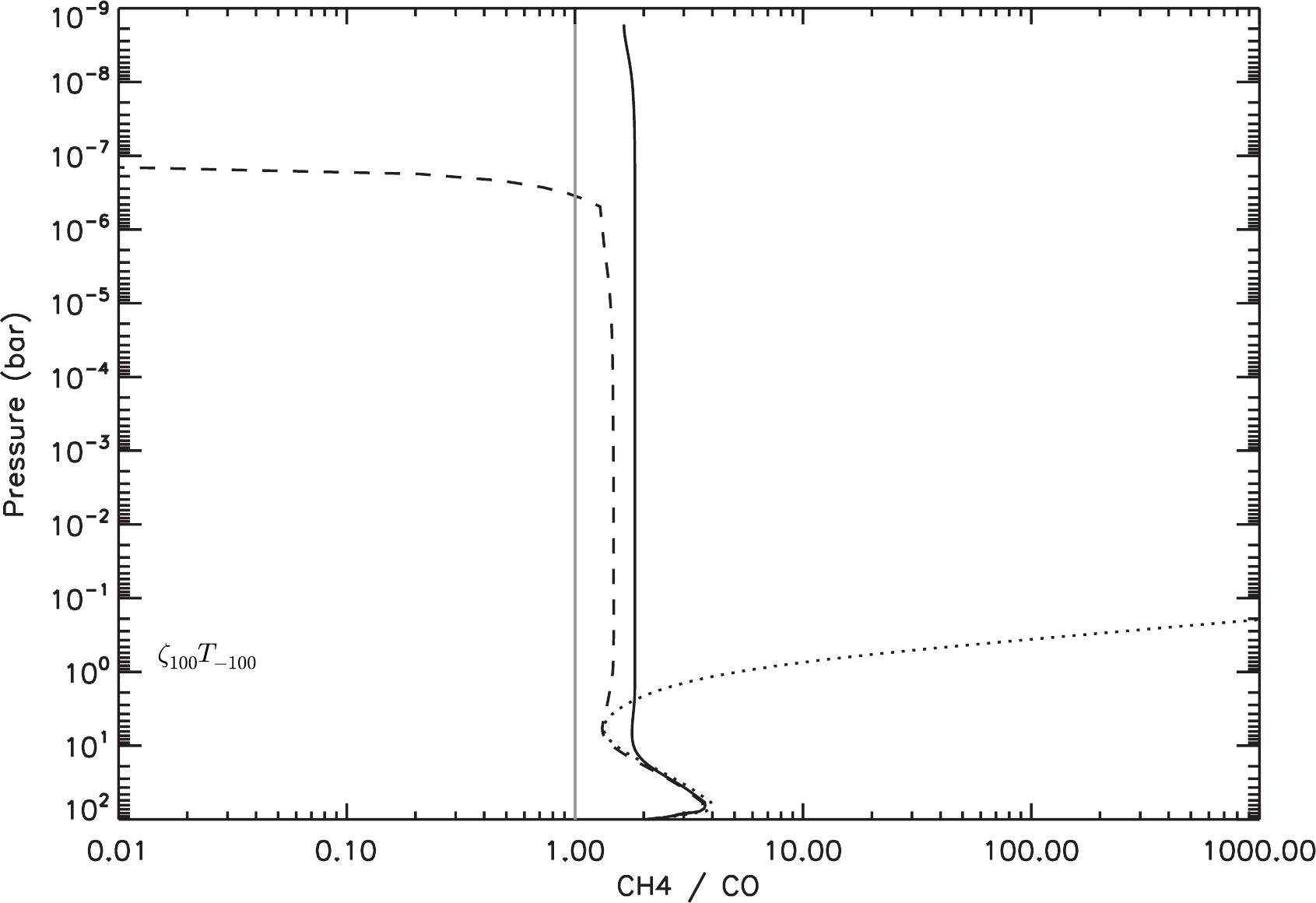}
\caption{Vertical abundances of CO and \ce{CH4} in eight selected models (left) and the corresponding value of the \ce{CH4}/CO ratio (right). Each line style correspond to a set of eddy diffusion coefficient and stellar UV flux (see legend in the left bottom panel and meaning of each symbol in Table~\ref{table:models_grid}). The \ce{CH4}/CO = 1 line is represented with a grey line.} \label{fig:abundances_CO_CH4}
\end{figure*}

\subsection{Combined effect of the parameters on the \ce{CH4}/CO abundance ratio}\label{sect:CH4_CO}

The main finding of this paper is that there exists a combined effect of the temperature, the vertical mixing, and the metallicity that can explain the \ce{CH4}/CO abundance ratio lower than unity found by observations in some atmospheres.
To clearly see the dependence of the \ce{CH4}/CO ratio with the parameters of our study, we plot for each $\zeta - T$ choice, the two more extreme cases, $K_{zz}^{\times10} F_{\lambda}^{\div 10}$ and $K_{zz}^{\div10} F_{\lambda}^{\times 10}$ (Fig.~\ref{fig:abundances_CO_CH4}). The two other cases are in between. 

In the cases of our low and standard metallicity, the \ce{CH4}/CO abundance ratio is always above 1 (for pressures higher than $10^{-2}$ mbar), whatever the choice of the parameters $T$, $K_{zz}$ and $F_{\lambda}$ (4 top panels on Fig.~\ref{fig:abundances_CO_CH4}). The maximum value reached is $\sim$200 between $10^{-3}$ and 1 bar (with the case $\zeta_{100}T_{-100}$). When the metallicity is increased up to $\zeta$ = 100, we find that the \ce{CH4}/CO abundance ratio may become lower than unity if the warm choice of temperature profile is adopted.
In the cases $\zeta_{100}T_{+100}$ (magenta curves on Fig.~\ref{fig:abundances_16models}), whatever the choice of $K_{zz}$ and $F_{\lambda}$, CO is clearly more abundant than \ce{CH4} (in the area $10^{-3}$--1 bar, \ce{CH4}/CO ranges between 0.04 and 0.06). We can see on Fig.~\ref{fig:abundances_CO_CH4} (bottom panels) that this result is determined by the thermodynamic equilibrium: because the temperature is high in the deep atmosphere, CO is thermochemically favoured over \ce{CH4}. Vertical mixing then makes the abundances of CO and CH$_4$ to quench in the vertical direction, so that CO remains more abundant than methane in all the upper atmosphere, despite thermochemical equilibrium predicts an inversion of C-bearing species between $\sim$0.7 and 100 mbar. Nevertheless, we find a \ce{CH4}/CO ratio higher than 1 for the cases $\zeta_{100}T_{-100}$ (blue curves on Fig.~\ref{fig:abundances_16models}). This indicates that not only a high metallicity is necessary to a \ce{CH4}/CO ratio under unity, but also a sufficiently high internal temperature.
 
These results show that it is possible for GJ~3470b, but also for the observed GJ~436b and, in a less extent because of its lower temperature, for GJ~1214b, to have a \ce{CH4}/CO ratio under unity. A very high metallicity, compared to the Sun, combined with a high temperature, may be the key to explain the observations of warm exoplanets, similar to GJ~3470b, which indicate that CO is more abundant than methane. Moreover, planets with a bulk composition similar to Neptune or Uranus are expecting to have such enrichment. The study shows also that getting back to the elementary abundances from observations is very difficult, and requires to know the temperature profile, the metallicity and the vertical mixing. The solution might be to use a self-consistent model taking into account and calculating simultaneously all these parameters.

\subsection{Synthetic spectra} \label{subsec:spectra}

\begin{figure*}[!ht]
\centering
\includegraphics[angle=0,width=0.98\textwidth]{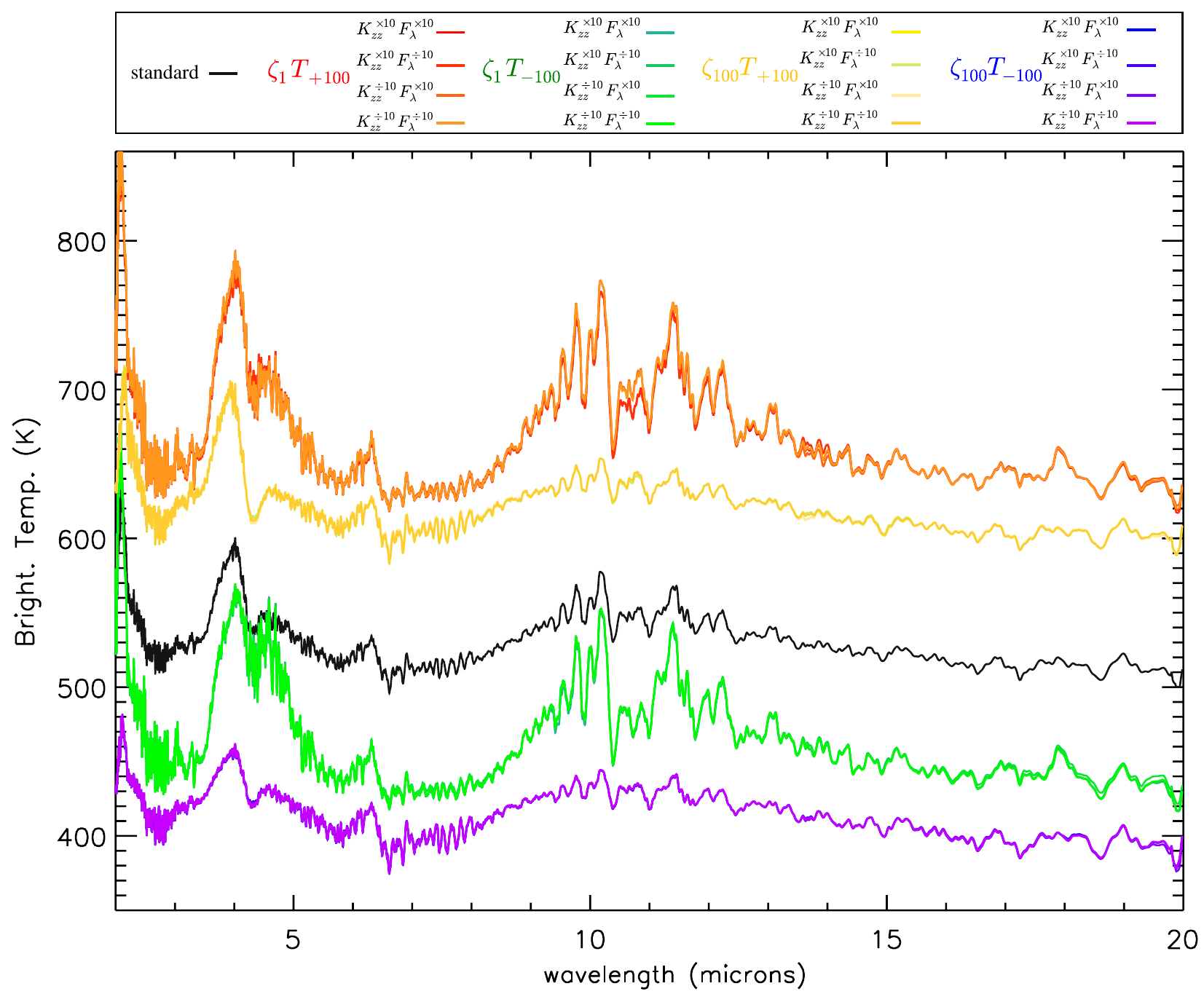}
\caption{Synthetic emission spectra of GJ~3470b corresponding to the grid of 16 models as well as the standard model. Each colour correspond to a set of metallicity and thermal profile. A colour gradient is then used to differentiate the eddy diffusion coefficients and stellar UV fluxes (see legend in the top panel and meaning of each symbol in Table~\ref{table:models_grid}). The standard values are shown in Figs.~\ref{fig:stellarspectrum} and \ref{fig:profile_pTKzz}. The standard metallicity is 10 $\times$ solar ($\zeta=10$).} \label{fig:spectra}
\end{figure*}
To determine if future observations of GJ~3470b could be used to constrain the values of the parameters that we varied in the previous section, we compute synthetic spectra for our 17 models (Fig.~\ref{fig:spectra}). Emission spectra are calculated using the line-by-line radiative transfer codes described in \cite{tin2005, tin2006}. Transmission spectra are calculated using the line-by-line radiative transfer codes described in \cite{hollis2013tau}. For both type of spectra, we used line lists from HITRAN \citep{roth2009, roth2010}, except for CO, \ce{CO2}, and \ce{CH4} for which we used HITEMP \citep{roth2010}. For \ce{H2O} we used the BT2 list \citep{barber2006}. We use the same NextGen stellar model as in Sect.~\ref{sec:stellar}.

We end up with five groups of spectra influenced by the thermal profile and the metallicity. The vertical mixing, as well as the UV flux used in the different models, have very little effect on spectra, which are dominated by the temperature and the metallicity of the atmosphere. Both for primary and secondary spectra, we notice that the reddish and greenish spectra exhibit broader variations than the two others, because they correspond to the low metallicity cases, so to atmospheres with smaller optical depth. This result has also been found by \cite{agu2013} and \cite{mos2013} for GJ~436b.

Thereby, for emission spectra, we see five levels of brightness temperature. The highest temperature corresponds to the case $\zeta_{1}T_{+100}$ (reddish curves), and the lowest to the case $\zeta_{100}T_{-100}$ (blueish curves). The standard model is logically in between the four other groups of spectra. With an identical thermal profile, the enhancement of metallicity (by a factor 100) leads to a lowering of the brightness temperature by $\sim$50K. The optical depth of the atmosphere increases together with the metallicity. Thus, the signal received during the secondary transit comes from higher in the atmosphere, so corresponds to lower temperatures, providing such differences of brightness temperature. Of course, in case of temperature inversion, one could have the opposite effect, and see higher brightness temperature with higher metallicities. Constraining the metallicity from the brightness temperature is made difficult by the strong dependency of this latter on the temperature, and therefore there is a degeneracy. Apart from the level of brightness temperature, the spectra are globally similar and exhibit the same features. Nevertheless, we can notice slight differences between the $\zeta_{1}$ and $\zeta_{100}$ spectra at two locations. First, around 4.5 $\mu$m, the absorption by \ce{CO2} and \ce{CO} are more defined in the $\zeta_{100}$ cases compared to the very close peaks characteristic of water absorption that we see around 4.5 $\mu$m in the $\zeta_{1}$ cases. Then, around 10 $\mu$m, the high peaks due to \ce{NH3} and \ce{H2O} are very strong features in the $\zeta_{1}$ spectra but are attenuated on the $\zeta_{100}$ spectra. This is due to the high abundance of \ce{CO2}, that absorb a lot from 9 $\mu$m (as much or even more than water), and thus contribute importantly to spectra. The variation of the other parameters (eddy diffusion coefficient and UV flux) has almost no impact on the emission spectra, except for the case $\zeta_{1}T_{+100}$ (reddish curves). Between 10 and 11 $\mu$m there are differences in the brightness temperature of about 10~K due to the change of ammonia abundance. Between 13.5 and 14 $\mu$m, the small variations of brightness temperature are attributed to HCN and \ce{NH3}, that both contribute strongly to the spectra in this wavelength region. Nevertheless, the differences from one spectra to another, due only to the change of eddy diffusion coefficient and UV flux, are very small, and probably not detectable with our current technologies (e.g. \citealt{ste2010}) do not obtained uncertainties lower than 20~K for GJ~436b with the \textit{Spitzer Space Telescope}).

\begin{figure*}[!ht]
\centering
\includegraphics[angle=0,width=0.98\textwidth]{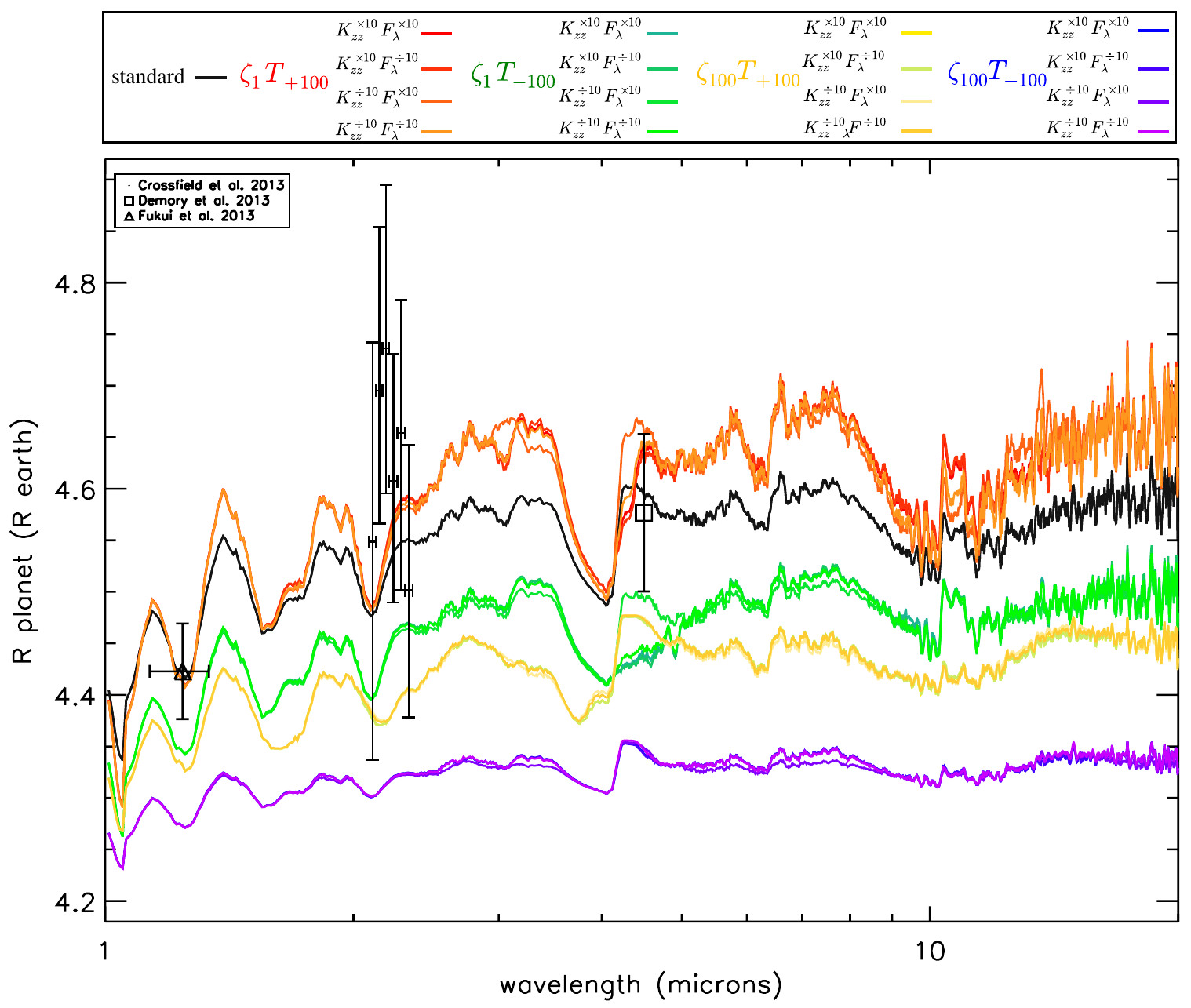}
\caption{Synthetic transmission spectra of GJ~3470b, in terms of apparent planetary radius, computed for all the 16 models of our grid as well as the standard model. Each colour corresponds to a set of metallicity and thermal profile. A colour gradient is then used to differentiate the eddy diffusion coefficients and stellar UV fluxes (see legend in the top panel and meaning of each symbol in Table~\ref{table:models_grid}). The standard values are shown in Figs.~\ref{fig:stellarspectrum} and \ref{fig:profile_pTKzz}. The standard metallicity is 10 $\times$ solar ($\zeta=10$).Observational data points (references in the legend) have also been plotted for comparison.} \label{fig:spectra_trans}
\end{figure*}

The transmission spectra are also separated depending on the temperature and the metallicity. Nevertheless, we notice that the greenish and yellowish spectra (respectively $\zeta_{1}T_{-100}$ and $\zeta_{100}T_{+100}$) are quite close and intersect between 4 and 5 $\mu$m although the chemical composition corresponding to these 8 cases are quite different. The apparent planetary radius found with our synthetic spectra goes from 4.25 to 4.75 R$_{\oplus}$. It is important to keep in mind that these numerical values depend on the choice of the radius of the planet at the 1 bar pressure level. Indeed, the observations give only the apparent radius of the planet and we cannot know to which pressure level it corresponds. Changing the radius at the 1 bar pressure level will translate the spectra vertically. What is important to study is the relative variation of spectra from one model to another. The radius at 1 bar is a parameter than can be adjusted to fit the observations. We decided to put the 1 bar pressure level at 4.28 R$_{\oplus}$, which corresponds to the minimum apparent radius observed \citep{dem2013}, slightly adjusted in order to fit the maximum of observational data points \citep{dem2013, fuk2013, crossfield2013} with the $\zeta_{1}T_{+100}$ and standard models. With a higher radius at 1 bar, the $\zeta_{1}T_{-100}$ and $\zeta_{100}T_{+100}$ (greenish and yellowish curves respectively) can also fit most of the observations. On the contrary, we see that the last case ($\zeta_{100}T_{-100}$) is too flat to be in the error bars. None of the models can perfectly match all the data points of \cite{crossfield2013}.\\
The higher radius is found with the model $\zeta_{1}T_{+100}$ (reddish curves), because the mean molecular weight is low (as opposed to the high metallicity cases), resulting in a higher atmospheric scale height, and thus in a higher radius. Compared to the yellowish curves, we see an increase up to 0.2 R$_{\oplus}$. On the opposite, the smaller radius is found  with the model $\zeta_{100}T_{-100}$ (blueish curves), because of the low atmospheric scale height due to the high mean molecular weight (4.1 g/mole).
We see that the 17 transmission spectra exhibit globally the same features, whatever the apparent planetary radius. Transmission spectra probe an upper part of the atmosphere (with a lower temperature), compared with emission spectra, and are thus more sensitive to UV photolysis, and vertical mixing. For one given $\zeta - T$ case, we can observe several variation on the spectral features. Around 3.3 $\mu$m and between 7 and 9 $\mu$m, we clearly see that the \ce{CH4} features change from one model to an other. Between 4 and 5 $\mu$m we see that the contribution of CO and \ce{CO2} evolve for the low metallicity cases only. It is consistent with the fact that the abundances of these two species almost don't change for the high metallicity cases (see Fig.~\ref{fig:abundances_16models}). Finally, the \ce{NH3} feature around 10 $\mu$m changes in the $\zeta_{1}T_{+100}$ case because the abundance of ammonia changes importantly with different $K_{zz}$ and $F_{\lambda}$ for this cases (see Fig.~\ref{fig:abundances_16models}). These variation are quite small (less than 0.1 R$_{\oplus}$) but could be detectable with current observational technologies. The \textit{Okayama Astrophysical Observatory} \citep{fuk2013} and the \textit{Hubble Space Telescope} \citep[][for the planet GJ~436b]{pont2009} for instance, are able to give error bars of only 0.1 R$_{\oplus}$. A detailed study like \cite{Tessenyi2013} applied to EChO and \cite{shabram2011} with JWST about the capacities of these future telescopes to differentiate our models is beyond the scope of this paper, but will be the subject of a follow-up study.

Although it is the only case with [CO]$>$[\ce{CH4}], the $\zeta_{100}T_{+100}$ spectra do not show strong features due to this different \ce{CH4}/CO ratio, except maybe the fact that the 3.3-to-4.7 $\mu$m radius ratio is very low on the transmission spectra. Observations at these wavelengths could thus prove useful as a diagnostic of the atmospheric carbon chemistry.

\section{Summary and discussion} \label{sec:summary}

We studied the atmospheric composition of GJ~3470b, a warm Neptune that is a promising target for spectral characterisation. In order to prepare and to predict these future observations, we explored the space of parameters that are uncertain (metallicity, vertical mixing, temperature of the atmosphere, and UV flux of the parent star) and computed 17 models. They allowed us to frame the different compositions that are possible for this planet. In most cases, the \ce{CH4}/CO ratio is above 1, although we found that under plausible conditions carbon monoxide becomes more abundant than methane. This can happen for the highest metallicity tested (100 $\times$ solar), which can be expected for planets with a similar bulk composition as Uranus and Neptune; we found that in this case, some models (with a high atmospheric temperature) lead to a \ce{CH4}/CO ratio under unity, down to a value of 0.04 in the $10^{-4}$--1~bar pressure range. We did not explore hotter temperature profiles because without increasing significantly the internal heat source (like what has been done by \citealt{agu2013} for GJ~436b) there is no reason to get such higher temperature for a given irradiation. It has already been shown with hot Jupiters that a higher temperature leads to a \ce{CH4}/CO ratio lower than 1 \citep[e.g.][]{mos2011, ven2012}. Moreover, our goal is not to map all the possible ranges of temperature, vertical mixing and metallicity that can produce a \ce{CH4}/CO ratio lower than unity, but to address how to get it.
Because of quite similar physical properties, this result can be extrapolated to other warm (sub-)Neptunes such as GJ~436b or GJ~1214b. Recently, a similar study has been carried out by \cite{mos2013}, who find also that a high metallicity could lead to a \ce{CH4}/CO ratio lower than 1 in GJ~436b. While the identification of the C-bearing species from observations is still under debate for these kind of planets \citep[][with GJ~436b]{ste2010,mad2011,bea2011}, these results show that even from a chemical model point of view the situation is not simple. \ce{CH4} may or may not be the major carbon reservoir, depending on both the metallicity, the temperature, and the vertical mixing. Indeed, we show in this paper that there is a combined effect of these parameters on the chemical composition of atmospheres. Because of quenching, the composition of the middle atmosphere can be affected by temperatures found much deeper than the observations. This carbon anomaly depends on the temperature contrast between the probed layers and the quenching level and on the efficiency of the vertical mixing.  At metallicity higher than 100 $\times$ solar, the vertical vertical mixing can propagate a CO/\ce{CH4} ratio above unity to the upper layers of the atmospheres. To retrieve the elemental abundances of such atmospheres, self-consistent models that couple all these influences are needed. Nevertheless, a very high metallicity ($\geq$100 times solar metallicity) seems to be a solution to explore to interpret future observations, as it is very likely for these atmospheres. The synthetic spectra we computed indicate that the brightness temperature as well as the transit depth vary significantly with the metallicity and the thermal profile, so future observations of GJ~3470b may be able to determine the metallicity and the temperature of this planet. Indeed, spectra corresponding to high metallicity models (100 $\times$ solar), because of the strong opacities, produce smaller features than low metallicity models (1 $\times$ solar). On primary transit, we found that the 3.3-to-4.7 $\mu$m ratio changes together with the CO/\ce{CH4} ratio. Observations at these wavelengths are a possible way to constrain this ratio.

\begin{acknowledgements}

O.V. acknowledges support from the KU Leuven IDO project IDO/10/2013 and from
the FWO Postdoctoral Fellowship programme. M.A., and F.S. acknowledge support from the European Research Council (ERC Grant 209622: E$_3$ARTHs). Computer time for this study was provided by the computing facilities MCIA (M\'esocentre de Calcul Intensif Aquitain) of the Universit\'e de Bordeaux and of the Universit\'e de Pau et des Pays de l'Adour.

\end{acknowledgements}

\bibliographystyle{aa}
\bibliography{bib_GJ3470}

\end{document}